\documentclass[prl,a4paper,twocolumn,showpacs]{revtex4}

\usepackage{graphicx}

\def\bN{\mbox{\bf{N}}}

\begin{document}

\title{Tracing ultrafast interatomic electronic decay processes in real time and space}

\author{Alexander I. \surname{Kuleff}}
\email[e-mail: ]{alexander.kuleff@pci.uni-heidelberg.de}
\author{Lorenz S. \surname{Cederbaum}}
\affiliation{Theoretische Chemie, PCI, Universit\"at Heidelberg\\
            Im Neuenheimer Feld 229, 69120 Heidelberg, Germany}
\date{\today}

\begin{abstract}
Tremendous advances in laser pump-probe techniques open the door for the observation in real time of ultrafast \textit{electronic} processes. Particularly attractive is the visualization of interatomic processes where one can follow the process of energy transfer from one atom to another. The interatomic Coulombic decay (ICD) provides such a process
which is abundant in nature. A multielectron wavepacket propagation method enables now to trace fully \textit{ab initio} the electron dynamics of the process in real time and in space taking into account \textit{all} electrons of the system and their correlations. The evolution of the electronic cloud throughout the ICD process in the rare gas cluster NeAr following Ne$2s$ ionization is computed and analyzed. The process takes place on a femtosecond timescale, and a surprisingly strong response is found at a much shorter attosecond timescale.
\end{abstract}


\maketitle

In recent years tremendous developments of laser pump-probe experimental techniques made possible a direct observation in real time of different kinds of ultrafast processes with sub-femtosecond resolution \cite{Drescher,Niikura,Niikura2,Niikura3}. This opened the door for the investigation of electronic processes in the attosecond/femtosecond timescale that take place before the nuclear dynamics comes into play. Processes like the rearrangement of the electronic system following an excitation of an inner-shell electron can now be traced in time and space and analyzed. These capabilities promise a revolution in our microscopic knowledge and understanding of matter.

The excitation of an electron from an atomic shell other than the outermost valence orbital creates a transient hole state. This hole is not stable and the system tends to minimize its energy by filling the vacancy with an electron from an outer shell, the excess binding energy being carried away either by an extreme UV or X-ray fluorescence photon, or transferred via Coulomb forces to another electron, which subsequently escapes from the atomic binding. When possible energetically, the latter non-radiative mechanism of de-excitation is extremely efficient in comparison to the competing photonemission. Indeed, the characteristic times are typically less than 100 femtoseconds (1 fs $=10^{-15}$ s) to even well less than a femtosecond entering the attosecond regime (1 as $=10^{-18}$ s), compared to radiative decay lifetimes which except for core levels of heavy elements, belong to the nanosecond range (1 ns $=10^{-9}$ s). 

The non-radiative decay processes can be divided into two major categories depending on whether the electrons involved in the process belong to the same or to different subunits of the system. The former are referred to as intra-atomic/molecular decay processes and the latter as inter-atomic/molecular decay processes. An example of intra-atomic decay is the well known Auger effect, following a core ionization of atoms, which besides its fundamental importance has a wide range of applications. Depending on the energy of the core hole and the strength of the electronic coupling, the Auger lifetimes range from few femtoseconds to few hundreds attoseconds (in the so-called super-Coster-Kronig transitions). That is why the observation in real time of the electron dynamics of such processes became possible  only recently. Few years ago, Drescher \textit{et al.} \cite{Drescher} using a sub-femtosecond X-ray pulse for excitation and a few-cycle light pulse for probing the emission traced the electron rearrangement during the M$_{4,5}$N$_1$N$_{2,3}$ Auger decay in Kr, with lifetime of the M-shell vacancy of about 8 fs, giving birth to time-resolved atomic inner-shell spectroscopy.

Contrary to core ionization, the ionization of inner-valence electrons usually produces ions in excited states lying energetically below the second ionization threshold thus making the slow radiative decay the only possible de-excitation mechanism as long as the resulting ions are isolated. It has been shown recently \cite{ICD1} that the situation is fundamentally different if the ions are embedded in an environment or have neighbors like in a cluster. Then, the possibility to distribute the positive charges between cluster subunits substantially lowers the double ionization threshold of the cluster compared to that of the isolated subunit giving rise to an inter-atomic/molecular decay mechanism, where the excess energy of the ion with the inner-valence hole is utilized to ionize a neutral neighbor. The process is ultrafast, in the femtosecond timescale, and the emitted electron has a kinetic energy of a few electronvolts. This process, called interatomic (or intermolecular) Coulombic decay (ICD), was predicted theoretically and shown to follow a general decay mechanism, taking place both in hydrogen bonded and in van der Waals clusters \cite{ICD1,SaCeICD_Auger,Vitali2}. Very recently, the theoretical findings have been confirmed in a series of spectacular experiments carried out by several groups \cite{Hergenhahn1,Doerner,Oehrwall,Hergenhahn_NeAr,Kyoshi}. Other experimental groups are engaged in the preparation of time-resolved ICD experiments using novel attosecond pulse techniques \cite{Leone,Kapteyn}.

The goal of the present work is to provide for the first time a theoretical description of an interatomic decay process in real time and space. Such a description requires accurate \textit{ab initio} calculations of the time evolution of the electronic cloud including explicitly the correlations among \textit{all} the electrons, i.e., it requires \textit{multielectron wavepacket dynamics}. Here, we present such a computational method and apply it for tracing in time and space the ICD process in NeAr following the $2s$ ionization of Ne. It should be noted that the method is equally suitable for calculating in real time all kinds of ultrafast electron relaxation processes following an ionization of a system. The technical details of the method are given elsewhere \cite{dprop}, where it is used to compute electron dynamics of \textit{non}-decaying states. Here, only the theoretical foundations of the method will be sketched as well as some subtleties concerning its application to decaying states. We stress that treating the dynamics of a decay process represents a much higher degree of complexity.

The starting point of our investigation is a neutral system $|\Psi_0\rangle$. The ionization of the system generates a non-stationary state $|\Phi_i\rangle$. The resulting hole charge then is traced in time and space, i.e., the time-dependent hole density is calculated. For convenience we assume that the initial state is created by the sudden removal of an electron. The \textit{hole density} of the ionized system is defined by 
\begin{equation}\label{eq1}
Q(\vec r,t):=\langle\Psi_0|\hat\rho(\vec r,t)|\Psi_0\rangle -
\langle\Phi_i|\hat\rho(\vec r,t)|\Phi_i\rangle =
\rho_0(\vec r)-\rho_i(\vec r,t),
\end{equation}
where $\hat\rho$ is the local density operator, and $|\Phi_i\rangle$ is the generated initial cationic state. The first term in Eq.~(\ref{eq1}) is the time-independent ground state density of the neutral system, $\rho_0$, and the second one, $\rho_i$, is the density of the cation which is time-dependent, since $|\Phi_i\rangle$ is not an eigenstate of the cation. The quantity $Q(\vec r,t)$ describes the density of the hole at position $\vec r$ and time $t$ and by construction is normalized at all times $t$. In the Heisenberg picture, the time-dependent part $\rho_i(\vec r,t)$ reads:
\begin{equation}\label{eq2}
\rho_i(\vec r,t)=\langle\Phi_i|e^{i\hat Ht}\hat\rho(\vec r,0)e^{-i\hat Ht}|\Phi_i\rangle = 
\langle\Phi_i(t)|\hat\rho(\vec r,0)|\Phi_i(t)\rangle,
\end{equation}
where $|\Phi_i(t)\rangle = e^{-(i/\hbar)\hat H t}|\Phi_i\rangle$ is the propagating multielectron wavepacket.

Using the standard representation of the density operator in a one-particle basis $\{\varphi_p(\vec r)\}$, often called orbitals, and occupation numbers $\{n_p\}$, Eq. (\ref{eq1}) can be rewritten in the following form
\begin{equation}\label{eq4}
Q(\vec r,t)=\sum_{pq}\varphi_p^\ast(\vec r)\varphi_q(\vec r) N_{pq}(t),
\end{equation}
where the matrix $\bN(t)=\{N_{pq}(t)\}$ with elements
\begin{equation}\label{eq5}
N_{pq}(t)=\delta_{pq}n_p - \sum_{M,N}\langle\Phi_i(t)|\tilde\Psi_M\rangle\rho_{MN}\langle\tilde\Psi_N|\Phi_i(t)\rangle
\end{equation}
is referred to as the hole density matrix. The second term of Eq. (\ref{eq5}) is obtained by inserting in Eq. (\ref{eq2}) a resolution of identity of a complete set of appropriate ionic eigenstates $|\tilde\Psi_M\rangle$. The matrix $\rho_{MN}$ is the representation of the density operator within this basis. 

Diagonalization of the matrix $\bN(t)$ for fixed time points $t$ leads to the following expression for the hole density
\begin{equation}\label{eq6}
Q(\vec r,t)=\sum_p|\tilde\varphi_p(\vec r,t)|^2\tilde n_p(t),
\end{equation}
where $\tilde\varphi_p(\vec r,t)$ are called \textit{natural charge orbitals}, and $\tilde n_p(t)$ are their \textit{hole occupation numbers}. The hole occupation number, $\tilde n_p(t)$, contains the information which part of the created hole charge is in the natural charge orbital $\tilde\varphi_p(\vec r,t)$ at time $t$. Because of the conservation of hole charge, one finds that $\sum_p\tilde n_p(t)=1$ at any time. The hole occupation numbers, together with the hole density, are central quantities in the observation and interpretation of the multielectron dynamics taking place after the removal of an electron.

For calculating the hole density matrix, Eq. (\ref{eq5}), we have used \textit{ab initio} methods only. The description of the non-stationary ionic state was done by means of the formalism of Green's functions, using the so-called \textit{algebraic diagrammatic construction} (ADC) scheme \cite{Ced_Encyc,non-Dyson}, and the direct time propagation of the electronic wavepacket was performed through the short iterative Lanczos technique (see, e.g., Ref. \cite{Leforestier}). For more details see Ref. \cite{dprop} and references therein. An important point should be addressed. Since the formalism is used for tracing the evolution of decaying states, i.e., a second hole is created in the system and an electron is ejected into the continuum, special care must be taken in constructing an appropriate basis set in order to have a good description of the continuum electron at least in some volume around the origin.

We have applied the above sketched methodology to describe in real time the interatomic decay of the Ne$2s$ vacancy in NeAr. The choice of the system is motivated by the availability of Ne$_n$Ar$_m$ clusters \cite{Hergenhahn_NeAr}, and by the fact that the lifetime of the Ne$2s$ hole (35fs, see below) is short compared to the nuclear motion (vibrational period of NeAr is 1.2 ps and the rotational period is 180 ps), justifying the use of clamped nuclei. Furthermore, the presentation simplifies by choosing a heteroatomic system where the initially ionized atom (Ne) is well distinguishable for its neighbor (Ar). The internuclear distance is taken to be 3.5~\AA, the equilibrium geometry of the NeAr cluster. To simulate an experiment one would have to compute the process at various internuclear distances and then to average over the weighted distribution of these distances. Such a procedure is too costy in view of the very large effort already invested in the present calculations. Nevertheless, such an averaging should only induce some smoothing of the curves presented below and a small shift of the predicted decay time. 

The electron dynamics calculations on NeAr have been performed using a combination of atomic and distributed Gaussian basis sets. The atomic basis set was chosen to be d-aug-cc-pVDZ on Ne and aug-cc-pVDZ on Ar. The distributed Gaussian basis consisted of (6$s$,4$p$,3$d$) Kaufmann-Baumeister-Jungen \cite{Kaufmann} diffuse functions centered between the Ne and Ar, and 36 $s$-functions placed on a grid on the $x$, $y$ and $z$ axes. The positions and the exponents of the latter are optimized to approximate up to 15~{\AA} from the system the radial part of the Coulomb $s$-wave with a kinetic energy of the ICD electron.

\begin{figure}[ht] 
\begin{center}
\includegraphics[angle=270,width=8.8cm]{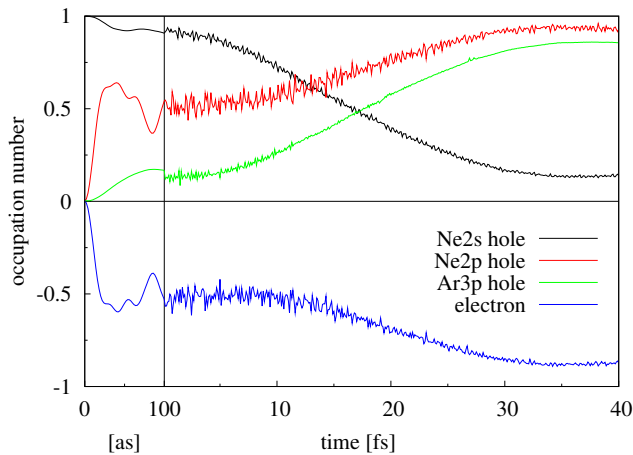}
\end{center}
\caption{(color online) Time-dependent hole occupation numbers of NeAr after Ne$2s$ ionization. Note the different timescale on the left part of the graphic. The initially ionized Ne$2s$ orbital, plotted in black, loses more than 80 \% of its hole charge in about 35 fs. In the same time, two other holes are opened -- one on Ne$2p$ (red curve), and one on Ar$3p$ (green curve) -- and more than 90 \% of an electron is created in the continuum (blue curve).}
\end{figure}

In Fig. 1 we show the results for the hole occupation numbers (see Eq. (\ref{eq6})) as a function of time for NeAr after sudden removal of a Ne$2s$-electron. This implies that $\tilde n_{Ne2s}=1$ at $t=0$  and all other hole occupation numbers are zero. As time proceeds, the initial hole (black curve) is gradually filled up, and two other holes are opened (red and green curves). This is accompanied by the creation of an electron (``negative'' hole) in the virtual orbital space (blue curve). For transparency, the hole occupations corresponding to the $p_x$, $p_y$ and $p_z$ components of the orbitals are grouped together, as well as all negative occupations contributing to the description of the continuum electron. The time evolution of the hole occupations reflects the timescale of the ICD process. After about 35 fs, the initial hole on Ne$2s$ is filled by an electron from the Ne$2p$ orbitals and an electron from the Ar$3p$ orbitals is ejected into the continuum, represented in our treatment by a vast number of virtual orbitals. The ICD lifetime is thus about 35 fs, in a very good agreement with the result obtained by elaborate \textit{ab initio} calculations of the decay width \cite{NeArSimona}. Several decay channels participate (see below) and, therefore, the shape of the Ne$2s$ hole occupation curve is not purely single-exponential, but rather a linear combination of several exponential functions. In principle, one may think of another process -- the hole on Ne$2s$ is filled by an electron from the Ar$3p$ and an electron from the Ne$2p$ is ejected. However, our numerical analysis of the propagating electronic wavepacket shows that the decay probability of this pathway is negligible being several orders of magnitude smaller than that of the ICD process.

It is worth mentioning that due to the finite basis set used, after some time the hole occupation numbers cease to reflect the physical reality when dealing with decaying states, since the so described continuum electron actually cannot leave the system. When the ICD electron reaches the spatial end of the basis set it can be reflected back yielding unphysical oscillations of the hole occupation numbers. In the present study such oscillations appear after about 50 fs. That is why, the size of the space covered by the basis set is of crucial importance for the proper description of decaying states. 

An interesting phenomenon is observed in the ultrashort timescale after the sudden ionization. On the left-hand side of Fig. 1 the first 100 as of the process are presented on an expanded scale. Besides the fast drop of the initial occupancy for about 50 as, shown to be universal for multielectron systems and related to the filling of the exchange-correlation hole \cite{JoergPRL}, one observes an extremely strong response of the system to the creation of the Ne$2s$ hole. In just 30 as, more than half of a full hole is opened on Ne$2p$ and more than half of an electron is already in the continuum. In such a short time the system is already ``prepared'' for the consecutive ICD process which is completed 35 fs later. The removal of a Ne$2s$-electron is seen to introduce an enormous disturbance of the electronic cloud yielding an extremely fast hole-particle excitation. We mention that at such short times a local ``violation'' of the energy conservation is possible. Indeed, following the Heisenberg uncertainty relation, $\Delta E\Delta t \sim \hbar$, one finds for times $\sim$ 30 as an energy dispersion $\Delta E\sim220$ eV! This is by far more than the energy involved in the whole ICD process.

\begin{figure}[ht] 
\begin{center}
\includegraphics[angle=270,width=9.5cm]{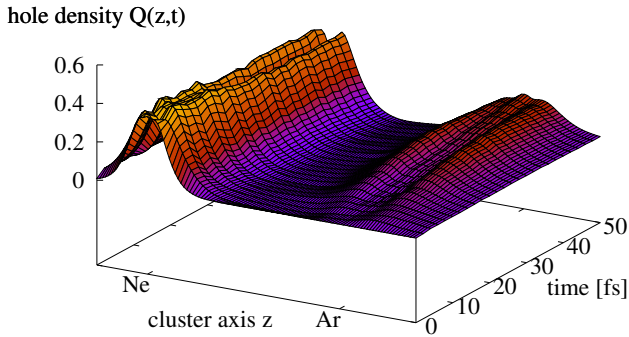}
\end{center}
\caption{(color online) Hole density $Q(z,t)$ plotted against the Ne-Ar axis $z$ as a function of time after Ne$2s$ ionization of NeAr. $Q(z,t)$ is obtained by integrating the total charge over the coordinates perpendicular to the $z$ axis. The initial hole density changes its character from an one- to a two-ridge surface, corresponding to the filling of the Ne$2s$-hole predominantly by a Ne$2p_z$-electron. In the same time a Ar$3p$-hole is created (right-hand side of the plot), which also has mainly $p_z$-character up to about 40 fs. One clearly sees that the dominant ICD-channel is Ne$^+(2s^{-1})$Ar$\to$Ne$^+(2p^{-1}_z)$Ar$^+(3p^{-1}_z)$.}
\end{figure}

More insight into the process is gained by inspecting the time evolution of the hole density, Eq. (\ref{eq6}). In Fig. 2 the charge $Q(z,t)$ obtained by integrating $Q(\vec r,t)$ over the coordinates perpendicular to the Ne-Ar axis ($z$ axis)  is plotted.  The displayed results support the conclusion that the dominant decay is to Ne$^+$($2p^{-1}_z$)Ar$^+$($3p^{-1}_z$); see also Ref. \cite{NeArSimona}. Indeed, the initial hole density (left-hand side of the surface in Fig. 2) changes its character from an one- to a two-ridge surface, corresponding to the filling of the initial Ne$2s$-hole predominantly by a Ne$2p_z$-electron. The second hole opened on Ar (right-hand side of the surface) displays predominantly $p_z$-character up to about 40 fs. The channel ending with Ar$^+(3p^{-1}_{x,y})$, seen as a tiny, steadily increasing ridge situated between the two wide humps on the right-hand side of the surface, becomes important after that time indicating a slower decay channel.

\begin{figure}[ht] 
\begin{center}
\includegraphics[width=8.2cm]{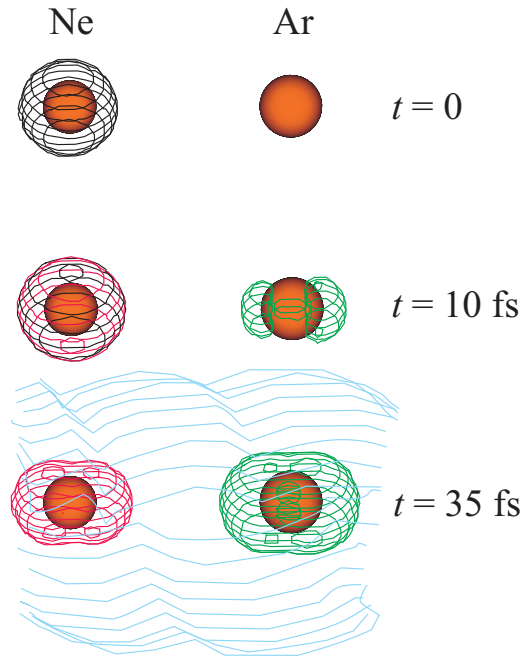}
\end{center}
\caption{(color online) The contribution of the holes ($\tilde n_p(t)\geq 0$ in Eq. (\ref{eq6})) to the hole density $Q(\vec r,t)$ at times 0, 10 and 35 fs. For completeness, at $t=35$ fs the electron density in the continuum ($\tilde n_p(t) < 0$ in Eq. (\ref{eq6})) is also given (light blue). The colors used correspond to those utilized in Fig. 1.}
\end{figure}

The surface in Fig. 2 represents the total charge density obtained by summing over both positive and negative occupancies, i.e., over hole and electron charges. Because of the mutual cancellations in some spatial regions, it is insightful to separate the hole and electron charges, i.e., to show separately the positive and negative occupancies. In Fig. 3 the so determined hole charge density $Q(\vec r,t)$ is shown in 3D for the three different times $t=0$, 10 and 35 fs. One clearly sees the stages of the ICD process. After 10 fs the initial $2s$ hole on Ne is already a mixture of $s$- and $p_z$-character, and at the same time the $p_z$-hole on Ar is already partly formed. After 35 fs the decay via the faster channel of the ICD process is essentially completed and we see Ne$2p_z$ and Ar$3p_z$ holes, while the Ar$3p_{x,y}$ orbitals have just started to contribute to the formation of the Ar$3p$-hole. For completeness, the electron density in the continuum at $t=35$ fs is also displayed in Fig. 3.

Let us summarize. A fully \textit{ab initio} methodology for calculating in real time ultrafast electron relaxation dynamics processes after ionization of a system is proposed. The method gives the possibility to trace in time and space the electron dynamics throughout different non-radiative decays taking into account \textit{all} electrons of the system and their correlations. We have presented the first study of the multielectron dynamics of the interatomic Coulombic decay in NeAr following Ne$2s$ ionization. The results disclose many microscopic details of the decay process in space as time proceeds. To  trace the dynamics of the decay directly in the time domain with attosecond resolution, a pump-probe experiment can be done in analogy to that described in Ref. \cite{Drescher}. We hope that the present research will stimulate future experimental and theoretical studies.

Financial support by the DFG is gratefully acknowledged.

\end{document}